\documentstyle[preprint,aps,epsfig]{revtex}
\begin{document}
\tighten
\draft
\title{First order optical potentials and 25 to 40 MeV 
proton elastic scattering.}

\author{P.~K.~Deb${}^1$, K.~Amos${}^1$, and S.~Karataglidis$^{2}$}
\address{$^1$School of Physics, University of Melbourne, 
Victoria 3010, Australia}
\address{$^2$Theory Division, Los Alamos National Laboratory, Los
Alamos, New Mexico, 87545}
\date{\today}
\preprint{LA-UR-00-2124}
\maketitle

\begin{abstract}
The differential cross sections and analyzing powers from the elastic
scattering of 25 and 40 MeV protons from many nuclei have been
studied.  Analyses have been made using a fully microscopic model of
proton-nucleus scattering seeking to establish a means appropriate for
use in analyses of radioactive beam scattering from hydrogen with ion
energies $25A$ and $40A$~MeV.
\end{abstract}

We present the results of analyses of elastic scattering of 25, 30,
and 40~MeV protons from nuclei made using coordinate space optical
potentials formed by folding complex energy dependent effective
two-nucleon ($NN$) interactions with ground state density matrices
given by shell model descriptions of the nuclei.  The interest to find
a credible prescription of the optical potentials at these energies
lies with current and future analyses of data from the scattering of
$25A$ and $40A$ MeV radioactive ions from hydrogen targets.  Such
experiments are being made at many facilities throughout the
world \cite{First}.  These optical potentials are required not only
for analyses of the elastic scattering cross sections but also to
define the distorted waves and the transition operator for use in
distorted wave approximation (DWA) analyses of the cross sections from
the inelastic excitation of the radioactive ions. Measurements and
subsequent analyses of such inelastic excitations are feasible and
have been made recently \cite{prl} for the excitation of the $2^+$
(1.8~MeV) state in $^6$He.

At the energies considered in the present work (25, 30 and 40 MeV),
collective structures in the response function of a nucleus may
contribute above any specific microscopic description based on an
effective $NN$ multiple scattering theory. For example, if the energy
is consistent with excitation of a giant resonance, virtual excitation
of that resonance could contribute to the scattering.  Indeed past
studies~\cite{reson} indicated that such virtual excitation of the
giant resonances gives energy-dependent signatures in cross sections.
Those effects however are of the order of 1~mb/sr at most and so are
evident basically only at large momentum transfers for elastic
scattering.  The usual (phenomenological) optical potential sufficed
to give the bulk of the (elastic) scattering results in those
studies~\cite{reson}. Hence, one may expect that a first-order
microscopic description of the optical potential, based on single-site
$NN$ scattering in medium, could still produce good agreement with
data taken for energies in the range 25 to 40~MeV.

Still, at these energies, the specific character of the target
response may be needed to specify appropriately the effective $NN$
interaction one should use with a folding prescription to define the
optical potential microscopically.  If so, the standard prescription
we have used to date to define the effective interactions may need
some modification. Calculations at these energies using that standard
prescription and comparison with data would calibrate any such
modifications required.  Of course, if the specific response function
effects in the definition of the effective $NN$ interaction are of
sufficient import, their omission should be evident in the comparisons
of current model results with data from light mass targets first, and
at 25 MeV in particular, given the excitation energies of the giant
resonances and their variation with target mass.  Therefore we have
analysed proton elastic scattering data taken at both 25 and 40 MeV
and from a number of nuclei in the mass range $A = 6$ to 208. The
coordinate space optical potentials we have used were obtained by
folding effective $NN$ interactions with one body density matrices
(OBDME) obtained from reasonable models of nuclear structure and of
single particle (SP) bound state wave functions of the targets. The
method used was that with which successful analyses of cross-section
and spin-dependent data from 65 and 200~MeV proton
scattering \cite{Ka9597,Do9597} have been made from many nuclei
ranging in mass from $^3$He to $^{238}$U.  As before, all details of
the effective interactions and structure are preset and no {\it a
posteriori} adjustment or simplifying approximation is made to the
complex non-local optical potentials that result from this process
which, hereafter, we term as $g$-folding.

The effective $NN$ interactions for 25 and 40 MeV incident protons are
a mix of central, two-body spin-orbit and tensor attributes each
having a form factor that is a sum of Yukawa functions~\cite{Do94} and
with complex, energy and density dependent strengths obtained by
accurately mapping the ($NN$) $g$ matrices of either the Paris $NN$
interaction~\cite{La80} or the Bonn-B potential~\cite{Ma87}.  Those
$g$ matrices, solutions of the Brueckner-Bethe-Goldstone equations,
were generated~\cite{Do94} for diverse nuclear matter densities as
linked to the Fermi momenta of infinite nuclear matter.  Note that the
energy and density dependences of the complex effective $NN$
interactions so formed have been crucial in forming the optical
potentials that yield good predictions at 65 and 200
MeV~\cite{Ka9597,Do9597}.  All details of the process and the
resultant optical potentials are given in depth in a recent
review~\cite{review}.

The $g$-folding approach has been used herein with the same two
effective $NN$ interactions defined above as input.  There are slight
differences between the two sets of $g$ matrices.  To assess these
effective interactions and the optical potentials formed with them,
each has been folded with the same structures, OBDME and single
nucleon bound state functions, that were used in the analyses of 65
and 200 MeV proton elastic scattering~\cite{Do9597} from each nucleus
considered.  All the results shown were obtained from calculations
made using the code DWBA98~\cite{Ra98}.  Thus the non-locality aspects
one finds with complete $g$-folding (coordinate space) optical
potentials have been evaluated and used without approximation.

\begin{figure}
\centering\epsfig{figure=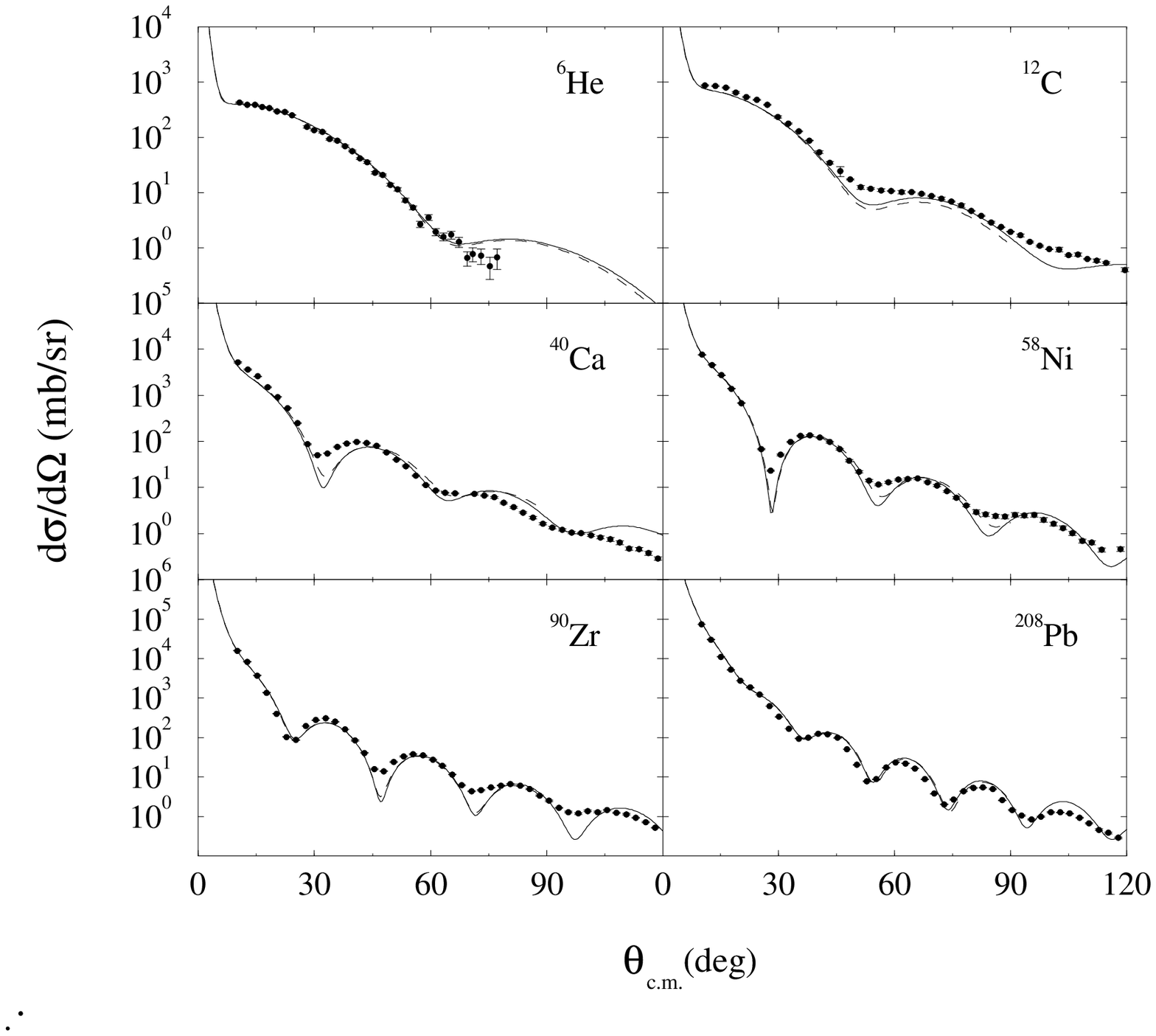,width=\linewidth,clip=}
\caption[]{The 40 MeV elastic proton scattering cross-section
data~\cite{First,Bl66} from a select set of nuclei compared with the
optical model calculations.}
\label{40xsecs}
\end{figure}

\begin{figure}
\centering\epsfig{figure=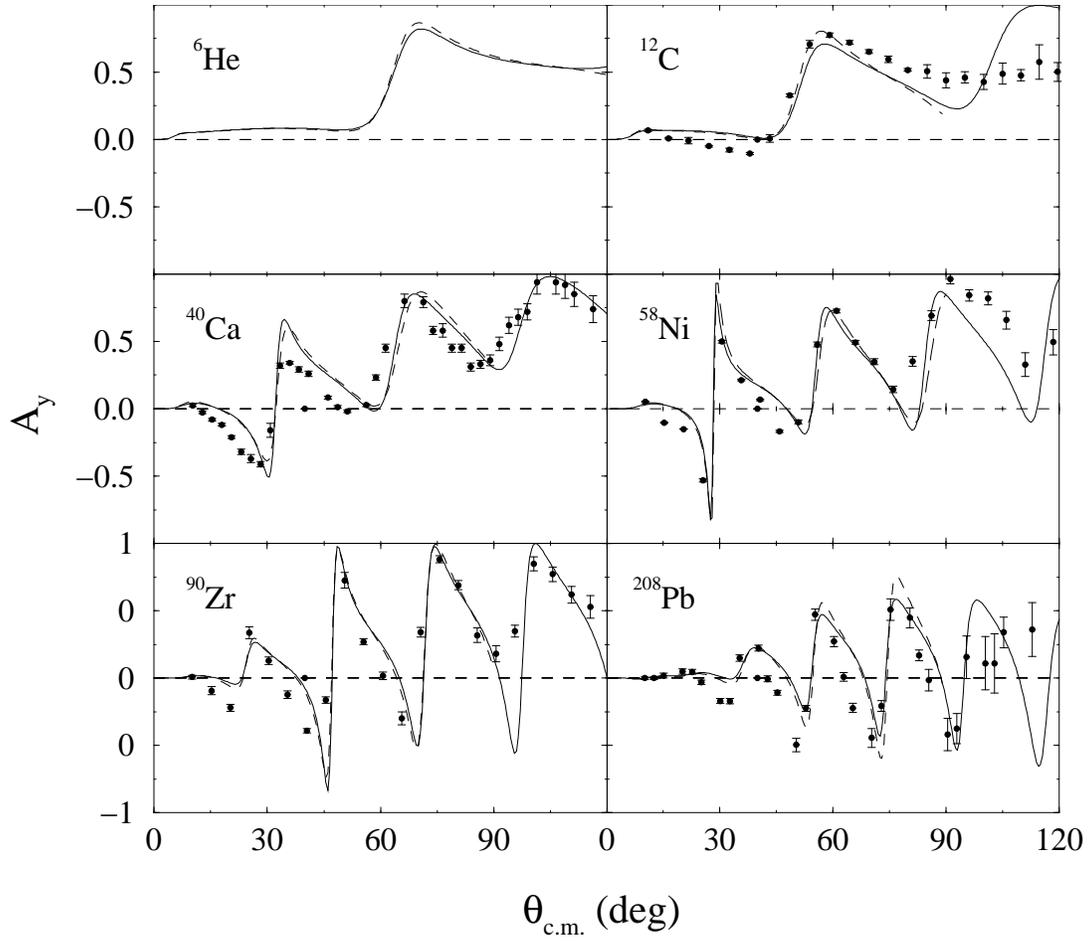,width=\linewidth,clip=}
\caption[]{The 40 MeV elastic proton scattering analyzing power data
from a select set of nuclei compared with the optical model
calculations.}
\label{40Ays}
\end{figure}

Use of the optical potentials so generated for 40 MeV proton
scattering gave the results shown in Figs.~\ref{40xsecs} and
\ref{40Ays}.  Therein the result for the exotic nucleus $^6$He is also
given as an example of use of the approach with radioactive ion -
hydrogen scattering.  For $^6$He, multi-$\hbar\omega$ space (no core)
calculations provided the OBDME as well as the harmonic oscillator
(HO) bound states to be used in the folding.  The Zheng $G$ matrix
elements~\cite{Zh95} were used in the $4\hbar\omega$ shell model
calculations \cite{Ka00} and a set formed by Navr\'atil and Barrett
were used for the $6\hbar\omega$ case \cite{Na98}.  For ${}^{12}$C,
$(0+2)\hbar\omega$ shell model wave functions were used while for the
heavier nuclei $0\hbar\omega$ or simple packed states were chosen.
The optical potentials were fixed therefore and single runs of the
scattering programs gave the results that are compared with data
obtained from Ref.~\cite{Bl66} for the stable nuclei and from
Ref.~\cite{First} with a radioactive beam of $40A$~MeV $^6$He incident
upon a hydrogen target.

In Fig.~\ref{40xsecs}, the calculated cross sections for 40 MeV proton
scattering are compared with data from a set of nuclei ranging in mass
from $^6$He to $^{208}$Pb.  The solid and long dashed curves in each
segment identify the $g$-folding results obtained by using the
effective interactions found from the Bonn-B and Paris potentials
respectively.  The differences in these calculated cross sections are
minor. These 40~MeV calculations agree with data almost as well as at
the higher energies~\cite{Do9597}, although most have more sharply
defined minima than is evident in the data.  By and large though, the
calculations are in good agreement with the shapes and magnitudes of
the cross-section data and now to 120$^\circ$ scattering.  For $^6$He,
the results found with the $4\hbar\omega$ and $6\hbar\omega$ models
are indistinguishable. Those structures, without adjustment to the SP
wave functions, do not give such an extended neutron distribution in
the ground state the ground state of $^6$He to classify $^6$He as a
neutron halo nucleus.  These elastic scattering data do not require
any such property of the target but the elastic data have not been
measured at momentum transfer values at which the non-halo versus halo
forms give noticeably different results.  However, the inelastic
scattering cross section with excitation of $^6$He to its first
excited 2$^+$ state does so~\cite{prl}.  The calculated cross sections
for the elastic scattering from $^{12}$C, $^{58}$Ni, and $^{90}$Zr are
quite reasonable in so far as the trend and magnitudes of the peaks
are concerned, with only too sharp minima being predicted.  The result
found for $^{40}$Ca is the worst and that for $^{208}$Pb the best.

In Fig.~\ref{40Ays}, the analyzing powers associated with 40 MeV
proton scattering from the same set of nuclei discussed above are
compared with data.  Again, the solid and long dashed curves portray
the $g$-folding results found with the Bonn-B and Paris interactions
as input.  The two forces give very similar results, and ones that
reflect the structure of the data quite reasonably.  Reflecting the
cross section predictions, that for $^{40}$Ca is the worst while that
for $^{208}$Pb is the best.

While the degree of replication of these 40~MeV data (differential
cross sections and analyzing powers) is not as good as found with the
higher energy data analyses, nevertheless all results are satisfactory
with no debilitating trend in quality of the fit with mass being
evident.  The study of 65~MeV scattering data revealed that medium
effects of the $NN$ interactions changed predictions by softening the
predicted cross section minima, and so these 40 MeV results may
indicate that an additional medium modification is needed in our
generated interaction.

\begin{figure}
\centering\epsfig{figure=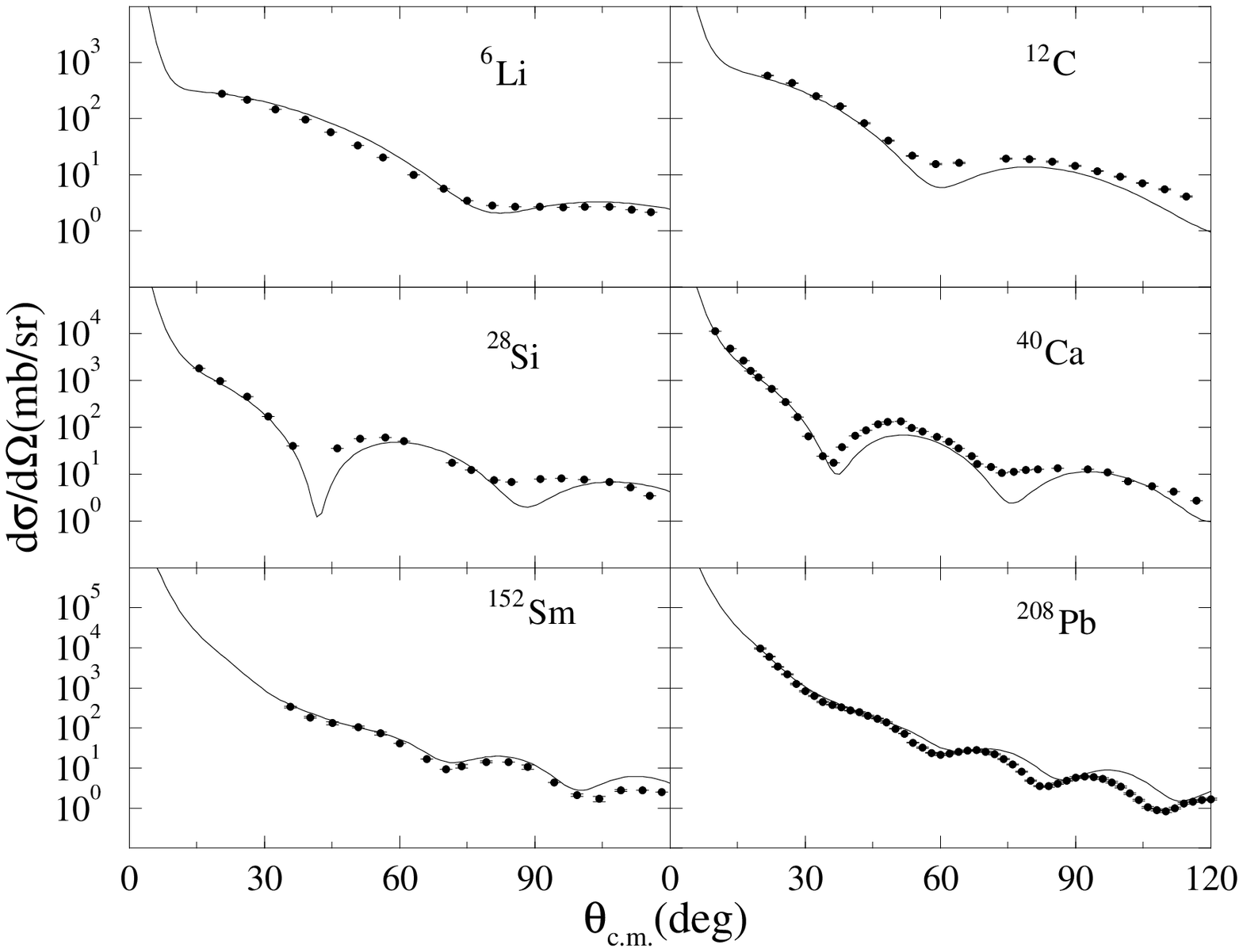,width=\linewidth,clip=}
\caption[]{The 25 MeV elastic proton scattering cross-section
data (Refs~\cite{Mu84} to \cite{Ri64})
 from a select set of nuclei compared with the
optical model calculations.}
\label{25xsecs}
\end{figure}

\begin{figure}
\centering\epsfig{figure=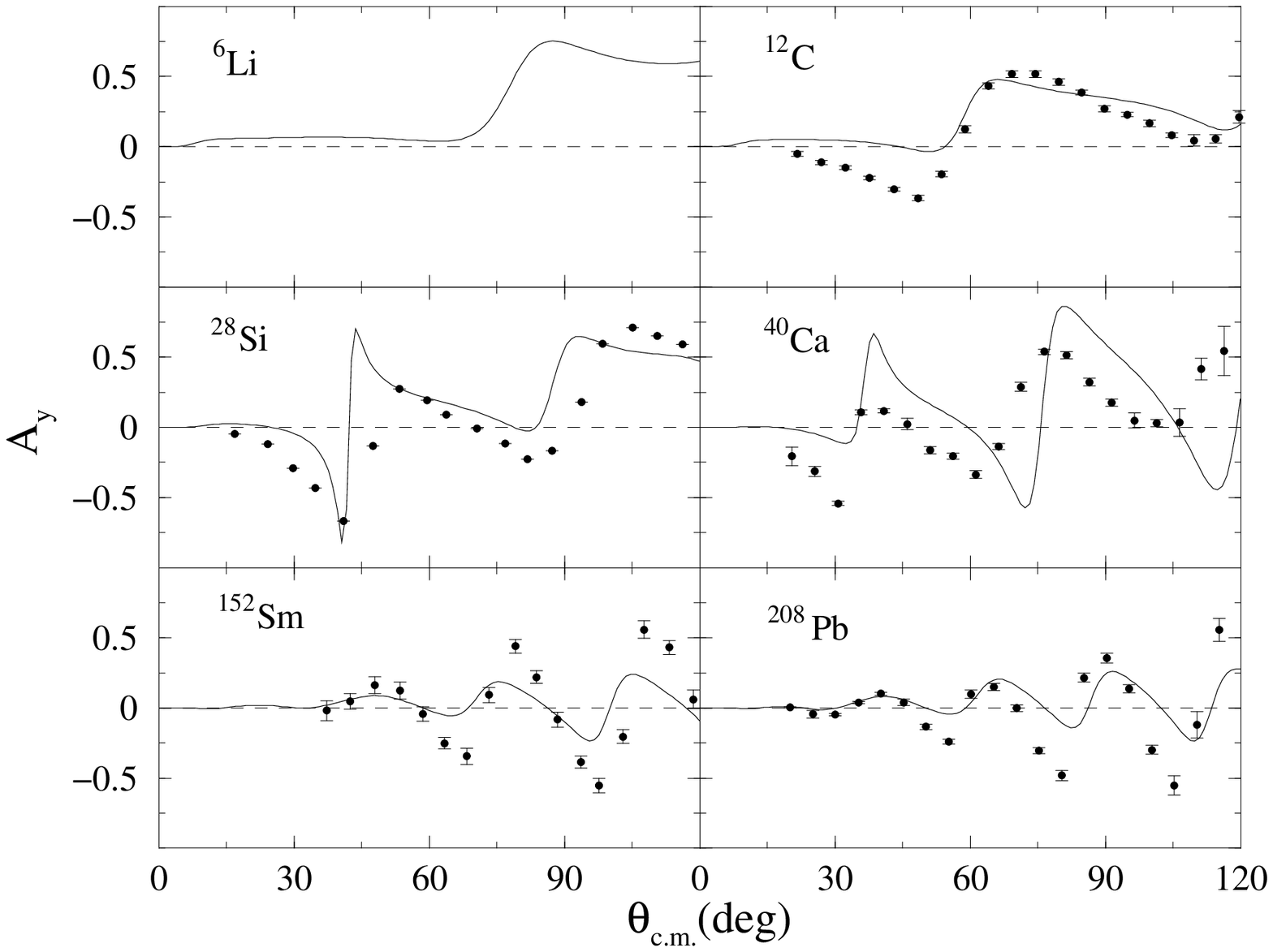,width=\linewidth,clip=}
\caption[]{The 25 MeV elastic proton scattering analyzing power data
(Refs.~\cite{La73,Ba71,Cr66,Cr64})
from a select set of nuclei compared with the optical model
calculations.}
\label{25Ays}
\end{figure}

The results of calculations made for 25~MeV proton scattering from the
nuclei, $^6$Li, $^{12}$C, $^{28}$Si, $^{40}$Ca, and $^{152}$Sm, and of
30~MeV proton scattering from $^{208}$Pb, are compared with
cross-section data in Fig.~\ref{25xsecs} and analyzing power data in
Fig.~\ref{25Ays}.  The differential cross section data shown were
measured at 25.9~MeV \cite{Mu84} for $^6$Li, at 24.0~MeV \cite{julich}
for $^{12}$C, at 25.0~MeV for $^{28}$Si \cite{La73} and $^{40}$Ca
\cite{Mcyy}, at 24.5~MeV \cite{Ba71} for $^{152}$Sm, and at 30.3~MeV
\cite{Ri64} for $^{208}$Pb.  The analyzing powers were measured at
24.1~MeV \cite{Cr66} from $^{12}$C, at 25.0~MeV \cite{La73} from
$^{28}$Si, at 29.0~MeV \cite{Cr64} from $^{40}$Ca, at 24.5~MeV
\cite{Ba71} from $^{152}$Sm, and at 29.0~MeV \cite{Cr64} from
$^{208}$Pb.  The comparisons of our calculated values with the
differential cross-section data are quite reasonable given that, at
this energy, higher order processes involving virtual excitation of
giant resonances~\cite{reson} may be expected to influence results for
$^{12}$C and $^{28}$Si in particular, and perhaps also for $^{40}$Ca.
The result with $^{208}$Pb may be sensitive to the choice of structure
and of the SP bound state functions in particular. A sensitivity to
the surface distribution of nucleons in heavy nuclei has been noted at
higher energies~\cite{review}.  As might be expected, the predicted
results of the analyzing powers are not in as good agreement with the
data.  Nevertheless, those calculated results follow the mass trend of
the measured values well enough to give credibility to the basic
attributes of our (first order) calculations and that our effective
$NN$ interactions at these energies are sensible.

The cross-section and analyzing power results obtained from the
coordinate space non-local optical potentials formed by $g$-folding at
40 MeV are in quite reasonable agreement with the data obtained with
targets of mass 6 to 208.  In general the cross-section predictions
give the magnitudes and trends of the peaks in the data but the minima
are too sharply defined.  The comparisons between the calculated
results and the data for 25 MeV proton elastic scattering remain
reasonable but the disparities are more pronounced than at higher
energies.  Nevertheless, the folded optical potentials remain a
reasonable first approximation, sufficiently so that the results may
still select between different structure inputs. Also the associated
distorted wave functions and effective interactions still should be
appropriate for use in distorted wave approximation analyses of
inelastic scattering from stable nuclei~\cite{Ka9597}, or of
radioactive beam ions, as well as of other reaction
calculations~\cite{Ri96}.

This work was partly supported by D.O.E. contract no. W-7405-ENG-36.


%

%

%

%

\end{document}